\documentclass[preprint]{emulateapj}

\slugcomment{Accepted for the Astrophysical Journal Letters}
\shortauthors{Zaritsky, et al.}
\shorttitle{}

\begin{document}
\title{Are Some Milky Way Globular Clusters Hosted by Undiscovered Galaxies?}
  
\author{Dennis Zaritsky\altaffilmark{1}, Denija Crnojevi\'c\altaffilmark{2}, and David J. Sand\altaffilmark{2}}
\altaffiltext{1}{Steward Observatory, University of Arizona, 933 North Cherry Avenue, Tucson, AZ 85721, USA; dennis.zaritsky@gmail.com}
\altaffiltext{2}{Physics Department, Texas Tech University, Box 41051, Lubbock, TX, 79409, USA}

\begin{abstract} 
The confirmation of a globular cluster (GC) in the recently discovered ultrafaint galaxy Eridanus II (Eri II) motivated us to examine the question posed in the title. After estimating the halo mass of Eri II using a published stellar mass - halo mass relation, the one GC in this galaxy supports extending the relationship between the number of GCs hosted by a galaxy and the galaxy's total mass about two orders of magnitude in stellar mass below the previous limit. For this empirically determined specific frequency of between 0.06 and 0.39 globular clusters per 10$^9$ $M_\odot$ of total mass, the surviving Milky Way (MW) subhalos with masses smaller than $10^{10} M_\odot$ could host as many as 5 to 31 GCs, broadly consistent with the actual population of outer halo MW GCs, although matching the radial distribution in detail remains a challenge. Using a subhalo mass function from published high resolution numerical simulations and a Poissonian model for populating those halos with the aforementioned empirically constrained frequency, we find that about 90\% of these GCs lie in lower-mass subhalos than that of Eri II. From what  we know about the stellar mass-halo mass function, the subhalo mass function, and the mass-normalized GC specific frequency, we conclude that some of the MW's outer halo GCs are likely to be hosted by undetected subhalos with extremely modest stellar populations.
\end{abstract}


\section{Introduction}
\label{sec:intro}

Based on the successful modeling of the internal dynamics and surface brightness profiles of globular clusters (GCs), it is generally thought that they are free of dark matter \citep{michie,king,conroy}. However,  
GCs are always found within a larger-scale dark matter halo, that of their host galaxy. 
In actuality, the number of GCs 
correlates with the host's total mass more strongly than it does with the host's luminosity or stellar mass \citep{blakeslee,mclaughlin,spitler,harris,hudson,harris15}. Does the relationship between the number of GCs and host's total mass extend into the subhalo regime? Could some of the Milky Way's GCs be hosted by yet undetected subhalos with extremely modest stellar populations? 

The suggestion that GCs could be hosted by subhalos originates with \cite{peebles}. 
The scenario is a natural outgrowth of the standard, dark-matter-dominated, hierarchical model \citep{moore}, and 
has more recently been adopted primarily as a means of understanding GC chemical anomalies \citep{bekki08, bekki}. Testing the scenario is difficult. Subhalos, as traced using their constituent stellar populations, become difficult to identify  in the ultrafaint galaxy regime because of the 
a precipitous decline in the stellar mass fraction with decreasing halo mass \citep[for a review see][]{kravtsov}. 
Among the known Milky Way (MW) low-mass subhalos, only the Fornax dwarf Spheroidal, which lies at the upper range of the MW dwarf satellite galaxy mass function, hosts multiple GCs. Tidally disrupted satellite galaxies, both of the MW and M31, may have hosted GCs \citep[][]{ibata,crane,martin,mackey,vel,li} but determining those subhalo masses is difficult. 

\cite{denija} confirm the existence of a GC in Eridanus II (Eri II) by matching color-magnitude diagrams (CMDs) for the GC and Eri II, thereby identifying the systems as coincident in 3-D, and measuring a size and luminosity for the GC that are within the ranges of MW outer halo GCs ($M_V = -3.5 \pm 0.6$ and $r_h = 13 \pm 1$ pc). The cluster CMD is consistent with that of Eri II, which is itself consistent with a 10 Gyr isochrone. The GC in Eri II \citep{koposov,denija} prompts a re-evaluation of the questions posed above and provides important empirical grounding for various arguments. First, Eri II and its GC 
extend the stellar mass, $M_*$, lower bound of MW subhalos that can host GCs by about two orders of magnitude. As such, the specific frequency, defined as the normalized number of GCs per 10$^9 M_\odot$ of stellar mass ($T_{N}$), is extremely high for this galaxy (Figure \ref{fig:spec_freq_mass}) and proves by example that subhalos of low total mass, $\sim 10^{9.2} M_\odot$ in this case, can host GCs. Second, the system constrains the GC specific frequency in low mass halos, which in turn enables an empirically-based, quantitative treatment of the problem. Third, the system proves by example that GCs can survive {as a recognizable GC} within such subhalos for a Hubble time.
In this {\sl Letter} we explore the implications of the Eri II GC using a simple, empirically-constrained model of the incidence of GCs within subhalos to determine whether we should expect subhalos to host some MW GCs.
We close by noting that \cite{denija} provide a strong upper limit on the Eri II H{\small I} content and do not find evidence for a young stellar population. As such, even though Eri II is currently at a large Galctocentric distance ($\sim$ 370 kpc), it may not be too dissimilar to other nearer MW satellites and is perhaps not on its initial infall.

\section{Results}
\label{sec:data}

Our data come from the study of Eri II \citep{bechtol,koposov,denija}, the compilation of Local Group dwarf properties \citep[][and references therein]{mcc}, 
and the study of GC populations in early-type \citep{z15} and late-type \citep{z16} galaxies in the S$^4$G survey \citep{sheth}. 

To calculate specific frequencies as defined most commonly in the literature, we require the number of GCs hosted by a galaxy and either that galaxy's luminosity, $L$, or $M_\star$. 
We focus on the $M_*$-normalized version of the specific frequency, $T_{ N}$. For Eri II, we 
use the published value of $M_*$ \cite[$83^{+17}_{-14}\times 10^3$ $M_\odot$;][]{bechtol}. To compare Fornax to Eri II, we calculate the implicit value of $M_*/L$ in the \cite{bechtol} study, 1.1 in solar units, and use that value to convert the available Fornax V-band magnitude \citep[$-13.4$ mag;][]{mcc} to $M_*$. To then place these estimates of $M_*$, which are based on a Chabrier initial mass function (IMF), on the same system as those from S$^4$G, which are based on a Salpeter IMF \citep{eskew}, we double the $M_*$ estimates for Eri II and Fornax. Our discussion depends only on relative values of $M_*$, so the choice of IMF is irrelevant to the degree that a universal IMF holds. 

\begin{figure}[ht]
\plotone{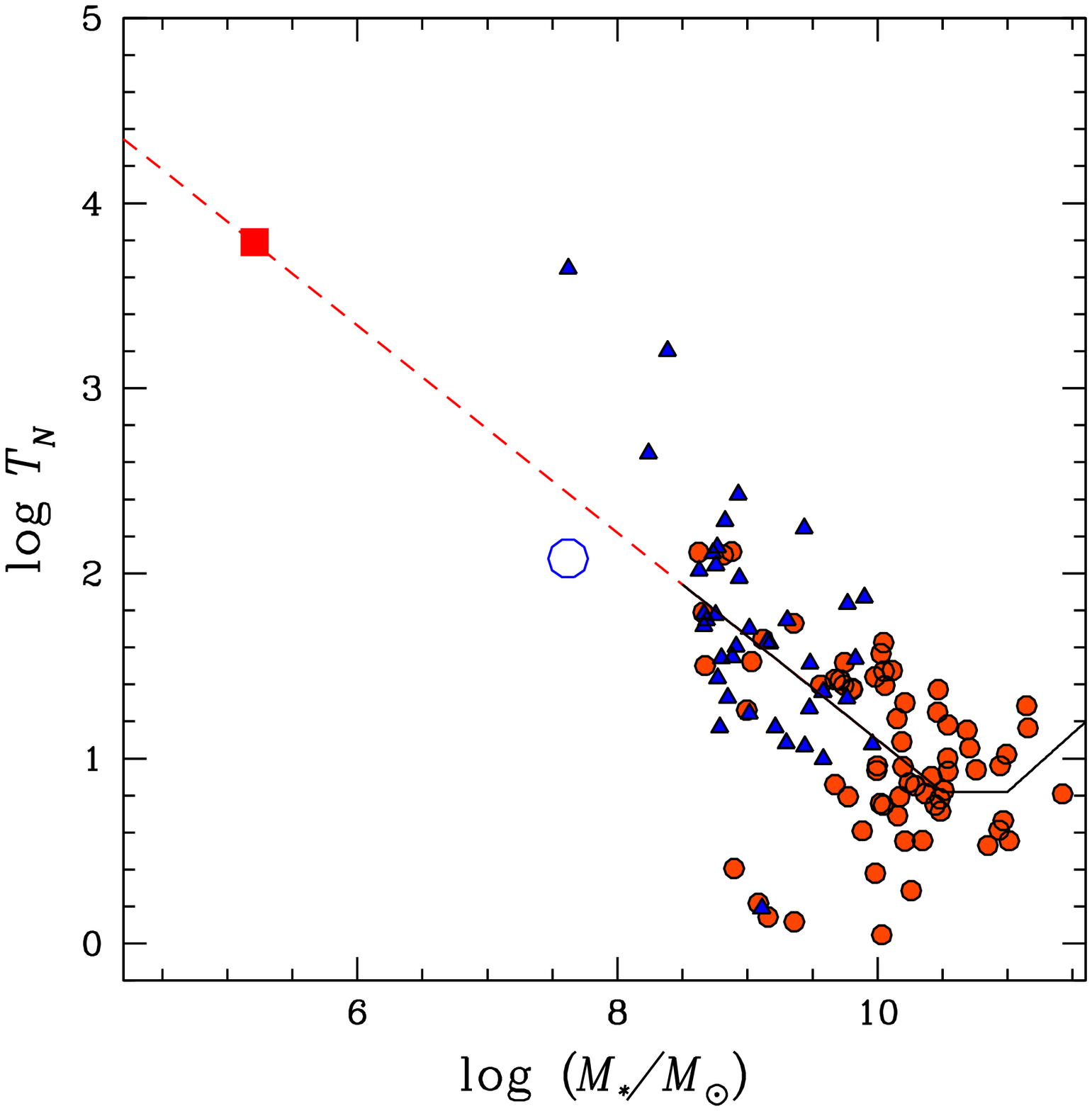}
\caption{The stellar-mass-normalized specific frequency of globular clusters, $T_{ N}$, vs. host-galaxy stellar mass. Scaling for $T_{ N}$ is number of GCs per 10$^9$ $M_\odot$. The red circles represent the early-type galaxies from S$^4$G \citep{z15}, the solid blue triangles represent the late-type galaxies from S$^4$G \citep{z16}, the solid line segments are the fit and extension presented in \cite{z16}, the dashed red line is a linear extrapolation of that fit to lower $M_\star$, the red filled square represents Eri II, and the blue open circle represents the Fornax dSph galaxy, which previously was the lowest- mass MW satellite known to host GCs. Uncertainties on the S$4$G individual points are similar to the scatter \citep{z16}, while the uncertainty in the Eri II and Fornax points are primarily in the stellar mass estimates. Errors in stellar masses tend to move points nearly along the dashed line.}
\label{fig:spec_freq_mass}
\end{figure}

In Figure \ref{fig:spec_freq_mass}, we reproduce the plot of $T_{ N}$ vs. host $M_*$ from \cite{z16} but expand the range to include Eri II. The Figure combines a) an existing datum for the Fornax dwarf Spheroidal, b) early- and late-type host galaxy results, c) a fit to those results for $8.5 < \log (M_*/M_\odot) < 10.5$ and extrapolation to higher $M_*$ based principally on the observations of \cite{peng} \citep[see][for details]{z16}), and d) the new results that include the Eri II datum and a linear extrapolation of the previously published fit extending to lower $M_*$. The extrapolation of the previous fit falls nearly exactly on the position of Eri II, suggesting that the GC populations in low-$M_*$ galaxies follow the trends established using high-$M_*$ ones.

\begin{figure}[ht]
\plotone{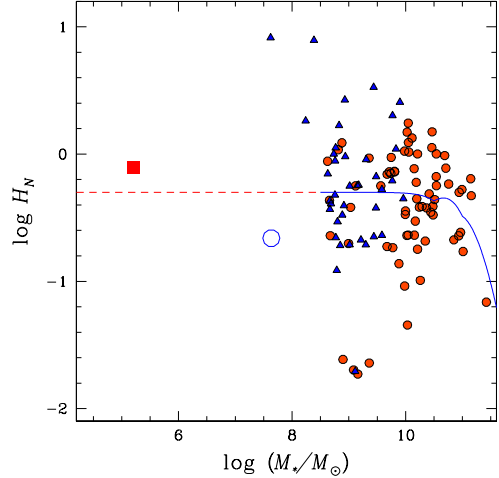}
\caption{
The halo-mass-normalized specific frequency of globular clusters, $H_{N}$, vs. host-galaxy stellar mass. Scaling for $H_{ N}$ is number of GCs per 10$^9$ $M_\odot$. Symbols are as shown in Figure \ref{fig:spec_freq_mass}.}
\label{fig:spec_freq_halo}
\end{figure}

Utilizing a large compilation of GC studies, \cite{harris} explored normalizing the GC number by host galaxy total mass rather than by stellar mass, following previous suggestions that either the total baryon or halo mass is the more physically-relevant parameter \citep{blakeslee,mclaughlin,spitler}. 
We adopt this idea in Figure \ref{fig:spec_freq_halo}, utilizing the functional form of the stellar mass - halo mass relation provided by \cite{behroozi} (correcting for the different adopted IMF and extending the relation beyond the mass range over which it is calibrated) to calculate the halo-mass-normalized GC specific frequency, which we name $H_{ N}$ and express in units of number per $10^9$ $M_\odot$, for all of the galaxies and the model shown in Figure  \ref{fig:spec_freq_mass}.

$H_{ N}$ is strikingly constant for $\log (M_\star/M_\odot) < 10$. Of course, this result depends on the nature of the stellar mass-halo mass function, which has sizable uncertainties even within the well-calibrated regime \citep{behroozi}, but we ourselves have not had any tuning freedom in reaching this result. As found for more massive galaxies \citep[e.g.][]{blakeslee,mclaughlin,harris} and in simulations \citep{kravtsov05}, the GC population closely tracks the overall halo mass. 

Although the two low-mass MW satellites that host at least one globular cluster fall tantalizingly close to the model extrapolation, there are 8 other known MW satellites that have stellar masses greater than that of Eri II and less than that of Fornax that do not host GCs. The lack of GCs in these systems does not necessarily invalidate the extrapolated model. 
First, because the number of GCs in any of these galaxies is small, only some subhalos, by chance, should be expected to host GCs. Second, additional physical effects might shape the GC population.  For example, some GCs might be tidally disrupted within their host subhalo or stripped away from their host subhalo by the MW's tidal field. Indeed the situation is likely to be quite complicated at small Galactocentric radii where both the clusters and subhalos will experience significant tidal forces and Galactic orbital times are shorter.
Therefore, we focus our discussion on the outer MW halo, but the appropriate choice for the lower radial bound to use in defining the outer halo is unclear. Given that the cluster NGC 1851 appears to lie within an extended stellar population of the type of interest here \citep{edo} and lies at a Galactocentric radius, $R_{GC}$, of 16.6 kpc, there could be hosted clusters at Galactocentric radii as small as 15 kpc.

We caution that detailed comparison between the data and models are premature. Our census of clusters similar to that in Eri II is woefully incomplete beyond $\sim$ 130 kpc \citep{koposov08}. More GCs with characteristics similar to that of the GC in Eri II ($M_V > -4$ and $r_{h} < 15$ pc) have been discovered in the past few years \citep[see][]{drlica} than are in the \cite{harris96} compilation. Our calculation is merely a plausibility argument and significant further work is necessary to transform this into a detailed comparison with discriminatory power.

\section{Discussion}

To enable progress,
we explore a simple Poissonian model. There are of course highly plausible reasons why Poisson statistics may be inappropriate, for example it may be that the existence of a first cluster naturally inhibits the formation of a second cluster. Nevertheless, given our ignorance, we adopt the simple Poissonian model and use the current data to set the one parameter that defines a Poisson distribution. Adopting a mean value of $H_N$, $\langle H_N \rangle$, we then realize the number of GCs in any subhalo by drawing from a Poisson distribution, where the mean number of GCs in a subhalo of mass $M_{sub}$ is simply $\langle H_N \rangle {\rm M}_{sub}$, where $M_{sub}$ is in units of $10^9$ $M_{\odot}$. 

\begin{figure}[ht]
\plotone{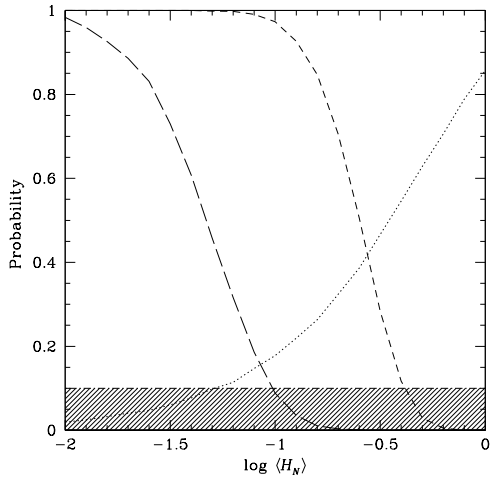}
\caption{Empirical constraints on $\langle H_N \rangle$. We calculate the probability of finding at least one cluster in a system with the mass of Eri II (dotted line), of finding 5 or fewer GCs in a system with the mass of Fornax (short dashed line), and of finding only 2 of 10 systems of comparable masses that host GCs (long dashed line). For each specific constraint, values of $\langle H_N \rangle$ where the curves lie within the shaded exclusion zone can be rejected with 90\% confidence.}
\label{fig:h}
\end{figure}

The key to this model lies in determining $\langle H_N \rangle$. We cannot simply adopt $H_N$ as measured for Eri II or Fornax because of the 8 other systems in this mass range that do not host GCs. Instead, we 
adopt the range of $\langle H_N \rangle$ 
for which we expect to find 1) a system with the halo mass of Eri II that has at least one GC, 2) a system like Fornax with 5 or fewer GCs, and 3) zero GCs in at least 8 of the 10 MW satellite galaxies in this mass range. 
This approach implicitly assumes that $\langle H_N \rangle$ is independent of halo mass for the halo mass range bracketed by Eri II and Fornax.
The results from Monte Carlo simulations are presented in Figure \ref{fig:h}. The Eri II and Fornax requirements provide upper and lower bounds on $\langle H_N \rangle$,  $-1.2 < \log \langle H_N \rangle < -0.4$ (90\% confidence limits). The additional constraint from the large fraction of satellites without GCs allows only for a small region around $\log \langle H_N \rangle=-1.1$. However, because we cannot be certain that physical processes have not removed or destroyed GCs in some of these satellites, we do not advocate for the calculated tight constraint around $ \log \langle H_N \rangle=-1.1$. Instead, we only explore the ramifications of $-1.2 < \log \langle H_N \rangle < -0.4$. 

\begin{figure}[ht]
\plotone{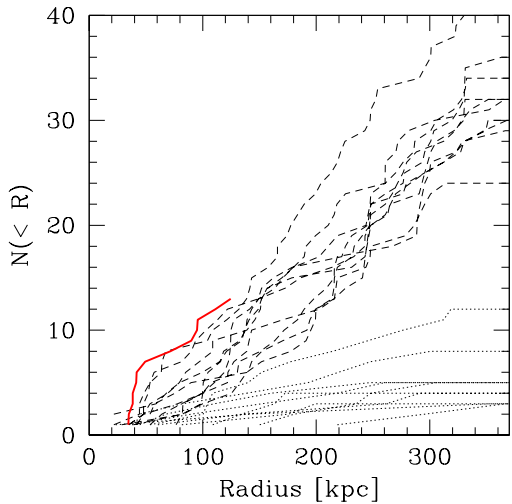}
\caption{Cumulative number of hosted GCs as a function of Galactocentric radius obtained by populating subhalos from published simulations \citep{madau,diemand} using the Poissonian model and our constrained range of $-1.2 < \log \langle {\rm H}_{\rm N} \rangle < -0.4$. The dashed lines represent 10 realizations using the upper end of the $\langle {\rm H}_{\rm N} \rangle$ range and the dotted lines the lower end of the range. For comparison, we overlay the cumulative distribution of outer halo GCs (radius $>$ 30 kpc) from the \cite{harris96} compilation as the solid red line.}
\label{fig:radial}
\end{figure}

Can some of the outer halo MW GCs lie within 
subhalos with masses as small or smaller than that of the known ultrafaint galaxies? 
The necessary calculation is fraught with various uncertainties and so the following is principally a plausibility argument. 
We begin by adopting the subhalo distribution from the Via Lactea simulations \citep{madau,diemand} for ${\rm M}_{sub}/{\rm M}_{\odot} < 10^{10}$ and a Galactocentric radius less than that of Eri II (370 kpc). 
We then use our Poisson model and random realizations to calculate the number of GCs found in subhalos of a given mass. From a set of realizations of this Poissonian model,
we find
that for the range of $-1.2 < \log \langle {\rm H}_{\rm N} \rangle < -0.4$, subhalos out to a radius of 370 kpc host between 5 and 31 GCs, respectively (Figure \ref{fig:radial}).

Straightforwardly, our simple model appears to
account for a significant fraction of the outer halo MW GC population. If we consider only the radial range over which our model predicts hosted GCs ($> 30$ kpc) out to the outermost cluster in the \cite{harris96} compilation ($\sim 125$ kpc), then that sample contains 13 clusters and compares favorably with the upper branch of the modeled 
distribution of clusters in Figure \ref{fig:radial}.
However, we caution against strong inferences from the current quantitative comparisons.
As the discovery of the GC in Eri II demonstrates, there may be GCs at large $R_{GC}$ that are yet to be discovered. 
The majority of GCs similar to the one in Eri II ($M_V > -4$ and $r_h < 15$ pc) have been discovered over the last few years and have a Galactocentric distance distribution that is markedly different than that of the previously known GCs \citep{drlica}. Specifically, only 10\% of the GCs in the \cite{harris96} compilation lie beyond 30 kpc, while 70\% (5/7) of the new GCs in this class do. The reason for this is evident in the completeness simulations of \cite{koposov08} when one considers that the newer clusters are significantly less luminous than most in the \cite{harris96} compilation.
In addition, various model parameters are poorly constrained. $\langle {\rm H}_{\rm N}\rangle$ varies by a factor of 6 and the subhalo bound mass fraction from simulations is not a detailed match to the MW, where the LMC alone accounts for a 25\% subhalo mass fraction \citep{penarrubia}.

About 90\% of the hosted GCs in our model are in subhalos with an $M_{sub}$ smaller than that of Eri II. As the image in \cite{denija} makes clear, the globular cluster is of higher surface brightness, and therefore easier to identify, than the underlying galaxy. It is not a great leap of imagination to conjure systems where the underlying stellar population is below the detection limit while the globular cluster is not. In fact, the stellar halo found surrounding NGC 1851 \citep{edo}, may be one example, the newly discovered Crater 2 dwarf and its potential globular cluster \citep{torr} another, and the large incidence of extratidal stars around inner halo MW GCs may point to the relics of such systems\citep{carballo}.
However, concluding that outer halo GCs could be hosted, is not the same as concluding that they are hosted. Deep observations will be necessary to determine if there are large-scale underlying stellar populations surrounding many outer halo GCs.

Another significant challenge to this model might appear to be the dynamical constraints on dark matter in GCs \citep[e.g.][]{conroy}. 
However, detailed modeling, such as that by \cite{conroy}, has been carried out on only a handful of GCs.  A similar handful of GCs, those in Fornax and Eri II, are almost certainly example of clusters that lie within a dark matter subhalo. We conclude that neither subset can be used as motivation for a blanket statement about the properties of all GCs.

Our scenario is quite different than that advocated by some \citep[e.g.][]{kravtsov05,kruijssen}, where the GCs form in high density, star forming disks at high redshifts and mergers redistribute the GCs into the galaxy halos. However, the models are not mutually exclusive. There are both kinematic and photometric distinctions within GC populations that hint at multiple formation paths \citep{brodie,dacosta}. 

\section{Summary}

The existence of one globular cluster (GC) in one of the ultrafaint MW satellites (Eri II) motivates our examination of how GCs populate low-mass subhalos. Using the \cite{behroozi} stellar mass - halo mass function, we estimate the halo mass of Eri II and the Fornax dSph and find that the total-mass-normalized frequency of GCs in these two galaxies is consistent with that measured for much more massive galaxies. This result suggests that the mean, mass normalized, GC frequency can be roughly extended to MW subhalos. For quantitative purposes, we use the observed GC incidence in Eri II and Fornax, and the lack of GCs in 8 other systems with comparable masses, to place bounds on the total-mass-normalized GC specific frequency, $-1.2 < \log \langle {\rm H}_{\rm N} \rangle < -0.4$, where the scaling is in terms of GCs per $10^9$ $M_\odot$. We calculate that subhalos with $ ({\rm M}_{sub} /{\rm M}_{\odot}) < 10^{10}$, which in simulations contain as much as 10\% of the MW halo mass \citep{madau}, may host 
somewhere between 5 and 31 GCs.
The order of magnitude agreement with the observed number of outer halo MW GCs demonstrates the plausibility of this model and suggests that some GCs may be hosted by yet undetected subhalos with extremely modest stellar populations.
The model further predicts that 90\% of the subhalos that host these GCs are of lower mass than that of Eri II, suggesting that host galaxy stellar populations will be difficult to detect. One such case may already be known \citep[NGC 1851;][]{edo} and examples where similar systems may have been tidally disturbed are increasingly common \citep{carballo,torr}. 

In closing, we encourage theorists to consider the effect that a hosting subhalo could have on a variety of long-standing questions regarding GCs and observers to revisit GCs when searching for the diffuse underlying stellar populations of ultrafaint galaxies.

\begin{acknowledgments}

DZ acknowledges support from NSF AST-1311326 and
DJS acknowledge support from NSF grants AST-1412504 and AST-1517649. We thank the referees for their comments, which helped us identify some outstanding issues and thereby helped clarify and strengthen our results and discussion.

\end{acknowledgments}

\end{document}